\documentstyle[12pt]{article}
\input epsf
\begin{document}
\title{
\hfill{\small TH-CERN/96-39}
\\
\hfill{\small TPBU-9-95}
\\
\hfill{\small hep-ph/9602313}
\\
\vspace*{0.5cm} \sc
A surprise in sum rules - modulating factors
\vspace*{0.3cm}}
\author{{\sc Alexander Moroz}${}^1$\thanks{On leave from Institute of
Physics, Na Slovance 2, CZ-180 40 Praha 8, Czech Republic}
\, and {\sc Jan Fischer}${}^{2*}$
\vspace*{0.3cm}}
\date{
\protect\normalsize
\it ${}^1$School of Physics and Space Research, University of 
Birmingham, Edgbaston, Birmingham B15 2TT, U. K.\\
${}^2$Theory Division, CERN, CH-1211 Geneva 23, Switzerland}
\maketitle
\begin{center}
{\large\sc abstract}
\end{center}
A generic physical situation is considered where Im $\Pi$,
the imaginary part of polarization operator (generalized 
susceptibility), can be measured on a  finite interval and the 
high frequency asymptotics (up to a few orders) of $\Pi$ can be 
calculated theoretically. In such a case, it is desirable to derive 
an equivalent form of the Kramers-Kronig dispersion relation,
the so-called sum rule, in which both the high-frequency part of 
Im $\Pi$ in the dispersion integral and the high-order contribution 
to $\Pi$ are suppressed. We provide a general framework for derivation
of such sum rules, without any recourse to an infinite-order differential operator. We derive sum rules for a wide set of weight functions and 
show that any departure from the $e^{-t}$ behaviour of the weight function 
in sum rules leads to modulating factors on the theoretical side of  
sum rules, providing its low frequency regularization.
We argue that by including  modulating factors one can extend the domain
of validity of sum rules further to an intermediate region of frequencies
and can account for ``bumps" which were observed numerically on
the phenomenological side of sum rules at ``intermediate'' frequencies.

\vspace*{0.2cm}

 
\vspace*{0.6cm}

\noindent February 1996\hfill

\thispagestyle{empty}
\baselineskip 20pt
\newpage
\setcounter{page}{1}
{\bf 1.} {\em Introduction}.-
Kramers-Kronig dispersion relations for
generalized susceptibility $\Pi$, such as \cite{Lans}
\begin{equation}
 \Pi(i\omega)= \frac{2}{\pi}
\int_0^\infty \frac{\mbox{Im}\,\Pi(s)}{s^2+\omega^2}\,sds,
\label{disr1}
\end{equation}
where $\omega$ is a frequency, arise in many branches of physics.
The reason is that
dispersion relations are  direct consequence
of the analyticity of generalized susceptibility
in the upper-half complex plane which,
in turn, is believed to be the consequence of causality.

Generalized susceptibility $\Pi$ determines the
response of the system to a perturbation.
Dispersion relation (\ref{disr1}) relates the value
of $\Pi$ at imaginary frequency to the integral over 
its imaginary part, Im $\Pi$. 
In the generic physical situation to be considered here,
Im $\Pi$ is determined experimentally on a finite interval
$(0,\omega_0)$ via directly measurable quantities such as
scattering cross sections.
On the other hand, we shall suppose that a few orders
of the asymptotic expansion of $\Pi$ for $\omega\gg 1$
 can be calculated theoretically.
An example is provided by quantum chromodynamics (QCD), where
the generalized susceptibility is the polarization
operator, defined via the current-current correlation function,
\begin{equation}
\Pi^{\mu\nu}(q) \equiv {\rm i} \int {\rm e}^{-{\rm i}qx} 
\langle 0|\, {\rm T} (j^{\mu}(x) j^{\nu}(0)) \, |0\rangle  
\,{\rm d}^4 x =(g^{\mu \nu}q^2 - q^{\mu}q^{\nu})\, \Pi (q^2),
\end{equation}
and  $q$  is a four-momentum.
The polarization operator satisfies the standard QCD dispersion
relation \cite{SVZ}
\begin{equation}
-\frac{d}{dQ^2} \Pi(Q^2) =
\frac{1}{\pi} \int_0^\infty
\frac{\mbox{Im}\,\Pi(s)}{(s+Q^2)^{2}}\, ds,
\label{disr}
\end{equation}
where $\Pi(Q^2)\equiv \Pi(-q^2)$.
The imaginary part, $\mbox{Im}\,\Pi$, is proportional to measurable 
cross sections such as that for $e^- e^+$ (electron-positron) 
annihilation into hadrons and, for sufficiently
small momenta, can be supplied from experiment. On contrary,
one can get, for $Q^2$ sufficiently large, the QCD representation
for the polarization operator as  (asymptotic) series,
\begin{equation}
Q^2\left(-\frac{d}{dQ^2}\right)
\Pi(Q^2)\sim a_0+ \sum_{n=2} \frac{a_n}{Q^{2n}}\cdot
\label{rex}
\end{equation}
Here,  the first term $a_0$ corresponds to asymptotic freedom
and the remaining coefficients $a_n$  ($n\geq 2$)
parametrize power corrections, describing long-distance effects. 
They are given as the product of a short-distance
term calculable within perturbation theory and a
corresponding (quark, gluon, etc.) condensate
which is unknown in theory \cite{SVZ}.
Therefore, in the generic physical situation,
it is desirable to find an  equivalent form of (\ref{disr1}),
a {\em sum rule},
in which both the high-energy part of $\mbox{Im}\,\Pi$
and the high-order contribution to $\Pi$ are suppressed,
 thereby reducing respective contributions  
of $\mbox{Im}\,\Pi$ and $\Pi$ for those frequencies and those 
orders for which our knowledge is incomplete.

The first step in this direction was performed
 by Shifman, Vainshtein, and Zakharov
(SVZ hereafter) \cite{SVZ}. By applying the formal
differential operator $\hat{L}_M$,
\begin{equation}
\hat{L}_M \equiv \lim_{
\stackrel{Q^2\rightarrow\infty,\, n\rightarrow\infty}
{\scriptstyle Q^2/n=M^2}} \frac{1}{(n-1)!} \left(Q^2\right)^n
\left(-\frac{d}{dQ^2}\right)^{n-1},
\label{lm-rel}
\end{equation}
on both sides of dispersion relation (\ref{disr})
they  derived the following equivalent
form of the dispersion relation:
\begin{equation}
\frac{1}{\pi}
\int_0^\infty e^{-s/Q^2}\mbox{Im}\,\Pi(s)\,ds \sim
Q^2 \sum_{n=0} \frac{a_n}{n! Q^{2n}},
\label{srf}
\end{equation}
 called the SVZ {\em sum rule}. In (\ref{srf}),
the contribution of $\mbox{Im}\,\Pi$  to the dispersion
integral  is reduced for energies $s\geq Q^2$
by an exponentially decreasing factor. On the right-hand side of
(\ref{srf}), the contribution of high orders of perturbation 
theory for $\Pi$ is suppressed due to the factor $n!$ in 
the denominator. Therefore, using sum rule 
(\ref{srf}), theoretical predictions and experimental results 
can be tested in a more efficient way than using 
original dispersion relation (\ref{disr1}).
In particular, in the case of QCD, one hopes that 
the validity of sum rule (\ref{srf}) can be extended to
intermediate energies and that the condensate values could be 
determined. A natural question arises whether one can provide a 
general framework for derivation of sum rules with a general weight 
function and specify their domain of validity. 

{\bf 2.} {\em Results}.-
To answer the above questions, we considered the class of
weight functions of the form
\begin{equation}
 \chi_{\alpha,\beta}= (1/\alpha)t^{(\beta/\alpha)-1}e^{-t^{1/\alpha}},
\label{chiab}
\end{equation}
where $0<\alpha\leq\beta$, and
\begin{equation}
\chi_\gamma (t)=t^\gamma e^{-e^t}
\label{chi0}
\end{equation}
with $\gamma\geq 0$.
Our choice (\ref{chiab}-\ref{chi0}) of weight functions covers
a wide range starting from monotonically decreasing to
``Gaussian-like" (cf. Ref. \cite{BL}) weight functions.

Our main result (see the next section) is
that a general sum rule with a weight function $\chi(t)$ can be
written as
\begin{equation}
\frac{1}{\pi}
\int_0^\infty \chi^{(1)}(s/Q^2)\,\mbox{Im}\,\Pi(s)\,ds \sim
Q^2 \sum_{n=0} (-1)^{(n-1)}\chi^{(n)}(\varepsilon_n /Q^2)
\frac{a_n}{n!Q^{2n}},
\label{gsr}
\end{equation}
where $\chi^{(n)}$ denotes the $n$th derivative of $\chi(t)$.
Constants $\varepsilon_n$ in (\ref{gsr}), $0\leq \varepsilon_n\ll Q^2$,  
are, in general,  different for different $n$.

We shall show that physically reasonable sum rules
arise only if $\chi(t)$ together with all its derivatives
is regular at the origin.
Then, in the region $\varepsilon_n\ll Q^2$,
\begin{equation}
\frac{1}{\pi}
\int_0^\infty \chi^{(1)}(s/Q^2)\,\mbox{Im}\,\Pi(s)\,ds \sim
Q^2 \sum_{n=0} (-1)^{(n-1)}\, \chi_n \,\frac{a_n}{Q^{2n}},
\label{fksum}
\end{equation}
where the coefficients $\chi_n$ are defined by
\begin{equation}
\chi(t)\equiv \sum_{n=0}^\infty \chi_n t^n.
\end{equation}
Sum rule (\ref{fksum}) includes all particular cases discussed
previously in Refs. \cite{SVZ,FK}.
Because weight functions are entire functions,
the coefficients $\chi_n$ as a function of $n$ decreases to zero 
faster than any polynomial in the limit $n\rightarrow \infty$.

Sum rule (\ref{gsr}) will be derived by employing  theory 
of summability methods,
without any recourse to  infinite-order
differential operator such as (\ref{lm-rel}).
This approach will also enable us to include into our consideration
the important case of weight function $\chi_\gamma$.
In contrast to $e^{-t}$ where frequency is required
to be imaginary, the weight function $\chi_\gamma$
allows one to consider dispersion relation (\ref{disr1})
for any frequency in the  upper-half complex plane.

In Sec. 4, we shall  show that unless
 $\chi(t)= e^{-t}$, one has to have $\varepsilon_n>0$. Thus
any departure from the $e^{-t}$ behaviour
of the  weight function in the sum rules leads
necessarily to $Q^2$-dependent factors
$\chi^{(n)}(\varepsilon_n /Q^2)$ on the theoretical side of
the sum rules, which we shall  call {\em modulating factors}.
Sum rules as a function of parameters
$\alpha$, $\beta$, and $\gamma$ [see Eqs. (\ref{chiab}) and (\ref{chi0})]
for a particular weight function are analyzed in Sec. 5.
 Domain of applicability of sum rules
(\ref{srf}) and (\ref{gsr}) and role of 
modulating factors are then discussed in Sec. 6.

{\bf 3.} {\em Derivation of general sum rules}.-
In what follows, the term ``summability method'' will stand for a
moment constant analytic (Borel-like) summability method
\cite{H,Mor}. In the theory of summability methods,
weight functions $\chi_{\alpha,\beta}$ and $\chi_\gamma$ are examples
of the generating function $\chi(t)$ which generates
the moments $\mu(n)$,
\begin{equation}
\mu(n)=\int_0^\infty \chi(t) t^n\,dt.
\label{genf}
\end{equation}
Any summability method is characterized
by its generating function which,
obviously, must decrease to zero  faster
than any power of $t$ in the limit  $t\rightarrow\infty$.
The properties of a summability method then depend
on the properties of the corresponding moment function $F(t)$,
determined in terms of moments $\mu(n)$ \cite{Fc},
\begin{equation}
F(t)\equiv \sum_{n=0}^\infty \frac{t^n}{\mu(n)}\cdot
\label{momf}
\end{equation}
In the case of $\chi_{\alpha,\beta}$, the moments are
$\mu(n)=\Gamma(\alpha n+\beta)$ and
the corresponding moment function
is known to be the Mittag-Leffler function \cite{H},
\begin{equation}
E_{\alpha,\beta}(z)=\sum_{n=0}^\infty\frac{z^n}{\Gamma(\alpha n+\beta)}\cdot
\label{mlf}
\end{equation}
In the case of $\chi_{\gamma}$, the moments $\mu(n)$ 
are growing roughly like
$(\ln\,n)^n$ and the corresponding moment function
is analyzed in \cite{Mor}.
Throughout the paper, we shall speak of Borel summability if and only
if $\chi(t)=e^{-t}$, in which case the moments
are $\mu(n)=n!$.
All other choices of $\chi(t)$  will be referred to as
the other summability methods.

For any weight function $\chi(t)$ discussed here,
the Cauchy kernel in dispersion integral (\ref{disr})
can be expressed as
\begin{equation}
\frac{1}{s+Q^2}=\int_0^\infty \chi(Q^2 t)\,F(-st)\,dt
=\int_0^\infty \chi(st)\,F(-Q^2t)\,dt,
\label{crep}
\end{equation}
where integrals in (\ref{crep}) are absolutely convergent \cite{H,Mor}.
Representation (\ref{crep}) is the key relation which
will allow us to represent the Cauchy kernel
in dispersion integral (\ref{disr}) as a
$\chi$-weighted integral and, at the same time, to find the
coresponding modification of the right-hand side of (\ref{disr}).

In the next, we shall assume dispersion integral
(\ref{disr}) to be  absolutely convergent \cite{Abs}.
Otherwise, one can always consider, instead of (\ref{disr}),
higher derivatives of the dispersion relation,
\begin{equation}
\frac{1}{n!} \left(-\frac{d}{dQ^2}\right)^{n} \Pi(Q^2)=
\frac{1}{\pi} \int_0^\infty
\frac{\mbox{Im}\,\Pi}{(s+Q^2)^{n+1}}\, ds.
\label{dcrep}
\end{equation}
 Relations of the type (\ref{dcrep})
do not bring any complications to our discussion,
since absolute convergence of the integral in (\ref{crep}) allows
us to differentiate within the sign of integration
and represent any power of the Cauchy kernel as
\begin{equation}
\frac{1}{(s+Q^2)^n}\equiv
\frac{1}{(n-1)!}
\left(-\frac{d}{ds}\right)^{n-1}\frac{1}{s+Q^2}=
\frac{1}{(n-1)!} \int_0^\infty F(-Q^2 t)
\left(-\frac{d}{ds}\right)^{n-1} \chi(s t)\,dt.
\label{qn}
\end{equation}
Again, for any $n$ the  integral in (\ref{qn}) converges
absolutely \cite{H,Mor}.
Then, by using relation (\ref{dcrep}) for $n=1$,
dispersion relation (\ref{disr}) can be written in the form
\begin{equation}
- \frac{1}{\pi}
\int_0^\infty \frac{1}{M^2}\, F(-Q^2/M^2)
\left(\int_0^\infty \chi^{(1)} (s/M^2)\,\mbox{Im}\,\Pi(s)\,ds\right)
d\frac{1}{M^2}=\sum_{n=0}\frac{a_n}{Q^{2(n+1)}}\cdot
\label{s-rule}
\end{equation}
Now,  the key problem is to find the function $G(1/M^2)$
such that  the right-hand side of
(\ref{s-rule}) can be  represented as an integral of the form
\begin{equation}
\sum_{n=0}\frac{a_n}{Q^{2(n+1)}} \equiv
\int_0^\infty\frac{1}{M^2}\, F(-Q^2/M^2) G(1/M^2)\,d\frac{1}{M^2}\cdot
\label{gin}
\end{equation}
In order to find $G(1/M^2)$, one again uses representation
(\ref{crep}) and (\ref{qn}) of the Cauchy kernel and its powers.
Since $Q^2$ is assumed to be sufficiently large,
one introduces only small error if $1/Q^{2n}$ is approximated
by $1/(Q^2+\varepsilon)^{n}$, where $0< \varepsilon\ll Q^2$.
Then, according to (\ref{qn}), general power $1/Q^{2n}$ can be
approximated as
\begin{equation}
\frac{1}{Q^{2n}}\sim \frac{1}{(Q^2+\varepsilon)^n} =
\frac{(-1)^{n-1}}{(n-1)!}
 \int_0^\infty F(-Q^2 t)\,\chi^{(n-1)} (\varepsilon t)\,t^{n-1}dt,
\label{qn1}
\end{equation}
which introduces an infinitesimally small correction to higher orders.
In what follows, we shall first  approximate
the right-hand side of (\ref{s-rule})
according to (\ref{qn1})  and  we shall discuss the
limit $\varepsilon\rightarrow 0$ afterwards.
Now, in the $\varepsilon$-approximation,
\begin{equation}
G(1/M^2) = M^2 \sum_{n=0} (-1)^{n}\chi^{(n)}(\varepsilon /M^2)
\frac{a_n}{n!M^{2n}},
\label{gnsum}
\end{equation}
and comparison with Eqs. (\ref{s-rule}) and (\ref{gin})
leads immediately to general sum rule (\ref{gsr}).

{\bf 4.} {\em Modulating factors}.-
A natural question arises what is the role
of these yet unspecified parameters $\varepsilon_n$.
Can one get rid of them?

In the case of the Borel method,  the generating function
$\chi(t)$ and the moment function $F(-t)$ are identical,
\begin{equation}
\chi(t) \equiv F(-t).
\label{sdual}
\end{equation}
We shall refer to this property  as
``{\em self-duality}''. It means that both $\chi(t)$ and $F(-t)$ are
exponentially decreasing and integrals in (\ref{s-rule}), (\ref{gin}),
and (\ref{qn1}) are absolutely convergent Laplace
integrals. In this case, obviously, $\varepsilon$ can be sent to zero in (\ref{qn1}),
\begin{equation}
G(1/M^2)=M^2 \sum_{n=0}\frac{a_n}{n! M^{2n}},
\end{equation}
and one recovers the SVZ sum rule (\ref{srf}).

For all other moment constant summability methods
discussed here, the ``self-duality'' (\ref{sdual}) is lost,
\begin{equation}
\chi(t)\neq F(-t).
\label{nos}
\end{equation}
Even more significant than the lack of ``self-duality"
property (\ref{sdual}) for the other summability methods is
that, in the case of the Borel method,  
$F(-t)$ ($=e^{-t}$) approaches zero faster than the inverse 
of any polynomial  in the limit $t\rightarrow\infty$,
while such inverse polynomial decrease is charactersitic 
for all other summability methods.
Indeed, for all other summability
methods, the asymptotic behaviour of $F(-t)$ in the limit 
$t\rightarrow\infty$
is characterized by an abrupt switch off from the {\em exponential}
decrease $e^{-t}$ to a  {\em universal polynomial} decrease \cite{Uni},
\begin{equation}
F(-t)\sim {\cal O}(t^{-1})\hspace*{3cm} (t\rightarrow \infty).
\label{unib}
\end{equation}
For example, in the case of
$\chi_{\alpha,\beta}(t)$
the moment function is $E_{\alpha,\beta}(t)$ [see (\ref{mlf})],
having  the asymptotic behaviour \cite{H}
\begin{equation}
E_{\alpha,\beta}(-t)\sim \sum_{n=1}^\infty 
\frac{(-t)^n}{\Gamma(\beta-\alpha n)}
\hspace*{2cm}(t\rightarrow \infty).
\label{mle}
\end{equation}
Unless both $\beta$ and $\alpha$ are integers
and $\beta\leq\alpha$, polynomial terms in the asymptotic expansion
(\ref{mle}) of
$E_{\alpha,\beta}(-t)$ are always present. Indeed, Eq.\,
(\ref{mle}) implies that $E_{\alpha,\beta}(-t)$
is exponentially decreasing if and only if for all $n$,
$\beta+\alpha n$ is a pole of the Euler gamma function $\Gamma(z)$.
For $\chi_\gamma (t)$ see \cite{Mor}.

The loss of the exponential decrease
of $F(-t)$ for all other summability methods implies
that the limit $\varepsilon\rightarrow 0$ cannot be taken.
Because of the universal behaviour (\ref{unib}) of $F(-t)$ at infinity,
\begin{equation}
\int_0^\infty F(-Q^2/M^2) \left(\frac{1}{M^2}\right)^n\,d\frac{1}{M^2}
\end{equation}
diverges for any $n\geq 0$ and
$G(1/M^2)$ {\em cannot} be pure polynomial. It must contain
modulating factors
which ensure convergence of (\ref{gin}) for small $M^2$.
Therefore, in general case, the sum rule corresponding
to the weight (generating) function $\chi(t)$ must be given
by relation (\ref{gsr}) with $0<\varepsilon_n\ll Q^2$.

{\bf 5.} {\em Analysis of sum rules for different 
weight functions}.-
Let us first consider $\chi_{\alpha,\beta}$ and  
denote $p\equiv\beta/\alpha-1$ and $q\equiv 1/\alpha$.
Then, in order to have meaningful sum rules,
$p$ and $q$ must be respectively nonnegative and positive integers. 
Otherwise, $t^n\chi_{\alpha,\beta}^{(n)}(t)\sim \chi_{\alpha,\beta}(t)$
in the limit $t\rightarrow 0$, resulting in the fact that all
orders in (\ref{rex}) will behave as ${\cal O}(1)$ in 
the limit $Q^2\rightarrow\infty$ on the theoretical side of sum 
rule (\ref{gsr}) and thus, they will merge to a single order 
and become {\em equivalent}.

If $p$ is a nonnegative integer and $q$ is a positive integer, 
then $\chi_{\alpha,\beta}$, including all its
derivatives, is regular at the origin.
Therefore, for $\varepsilon_n\ll Q^2$,  modulating factors 
 $\chi_{\alpha,\beta}^{(n)}(\varepsilon_n/Q^2)$
in (\ref{gsr}) can be approximated by their values at the origin
and one finds relation (\ref{fksum}) 
which, upon appropriate choice of the weight function, 
gives  the SVZ sum rule (\ref{srf}) and its generalizations
discussed in \cite{FK}.

To discuss particular cases, one uses 
\begin{equation}
\chi^{(n)}(0) = n!\chi_n=
(-1)^{l} q\,\frac{(p+lq)!}{l!} \delta_{n,p+lq}
=(-1)^{l}q\, \frac{n!}{l!} \delta_{n,p+lq}.
\label{derf}
\end{equation}
Therefore, unless $n= p+ lq$, where $l$ is a nonnegative integer,
$\chi_{\alpha,\beta}^{(n)}(0)=0$.
Relation (\ref{derf}) implies that
the sum rule with such  weight function
is (i) {\em selective}: it only selects the orders
$n=p+ lq$ of the asymptotic expansion (\ref{rex}),
(ii) the contribution of the $(p+lq)$th order in (\ref{rex}) 
is then reduced by the factor $1/l!$ on the theoretical side 
of the sum rule.
For example, in the case of $\chi(t)=2 e^{-t^2}$ only
even terms of the asymptotic expansion (\ref{rex})
enter the sum rule (\ref{fksum}).
Thus, such selective sum rules can be particularly useful in the case 
when some symmetries of the problem ensure that only 
$(p+lq)$th orders of the asymptotic expansion (\ref{rex}) are nonzero.

In the case of $\chi_\gamma$,
similar argument shows that it is useful to have $\gamma\neq 0$
only if the first $\gamma$ orders of the
asymptotic expansion (\ref{rex}) are identically zero.
Otherwise, for $e^{-e^t}$, one has
\begin{equation}
\frac{d^n}{dt^n}\,\exp\left( -e^t\right)\equiv
n!\bar{\chi}_n=\sum_{k=0}^\infty \frac{(-1)^k}{k!}k^n.
\label{chi0d}
\end{equation}
Estimates, using Euler-Maclaurin sum formula \cite{F}, give
$|\bar{\chi}_n|\sim (\ln n)^n/n!$ in the limit $n\rightarrow\infty$.

{\bf 6.} {\em Domain of validity of sum rules
and physical significance of modulating factors}.-
Sum rules (\ref{srf}) and (\ref{gsr}) are derived  for $Q^2$ 
sufficiently large. However, the 
meaning of ``sufficiently large'' must be decided separately in 
each particular case.
Since, eventually, one wants to make some practical use
of sum rules, one is interested in some intermediate region
of energies (or frequencies). The best possible situation
occurs if there is some nonempty intersection 
(''fiducial region'') of respective
intervals on which Im$\Pi$ and $\Pi$ are known.
Note that if the intersection is empty, one may gain by
repeating all our calculations with $\chi(t)$ replaced by
$\chi_A(t)=\chi(At)$, by introducing a scaling variable
$A$. This scaling leads to $F(t)$ replaced by
$F_A(t)=F(At)$ and $n!$ factor in (\ref{gnsum}) replaced by
$n!/A^n$. The corresponding sum rule is then
\begin{equation}
\frac{1}{\pi}
\int_0^\infty \chi^{(1)}(sA/Q^2)\,\Pi(s)\,ds \sim
Q^2 \sum_{n=0} (-1)^{(n-1)}\chi^{(n)}(\varepsilon_n A/Q^2)
a_n\,\frac{A^n}{n!Q^{2n}}\cdot
\label{gasum}
\end{equation}
For $A>1$, one gains on the experimental side of sum rule:
the interval on which $\mbox{Im}\,\Pi$
must be accurately determined is scaled down from interval
$(0,Q^2)$ onto interval $(0,Q^2/A)$.
However,  one lost on the theoretical side of 
the sum rule: (i) damping of higher order terms is slower, 
(ii) since the factor $A^n/n!$ has its maximum at $n=A$, 
terms in the expansion (\ref{rex}) at the order $n\approx A$ enter
the sum rule with relatively the highest weight.
Therefore, although the scale 
transformation is worth considering, 
it is not at all obvious whether one can extend
the validity of sum rules to the intermediate region.

We argue that modulating factors play important
role in the extension of validity of sum rules to the
intermediate region. The argument is as follows: by comparing 
the behaviour of both sides in (\ref{srf}) in the limit 
$Q^2\rightarrow 0$ one finds that the left-hand side goes to zero 
while the right-hand side diverges. 
On the other hand, if modulating factors are present then, 
in contrast to (\ref{srf}), both sides of (\ref{gsr}) tend to zero 
in the limit $Q^2\rightarrow 0$. Modulating factors
can be viewed as the low energy regularization of sum rules.
The mismatch between the vanishing rates  of 
the two sides of (\ref{gsr}) in this limit (the left-hand side tends 
to zero polynomially while the right-hand side tends to zero
faster than any power of $Q^2$) is not too important: 
nobody expects  sum rule (\ref{gsr}) to be valid for $Q^2$ too small.
Nevertheless, the very fact that, in the presence of modulating factors, 
both sides of (\ref{gsr}) tend to zero in the limit $Q^2\rightarrow 0$, 
represents substantial improvement over the SVZ sum rule. 

\begin{figure}
\centerline{\epsfxsize=8cm\epsfbox{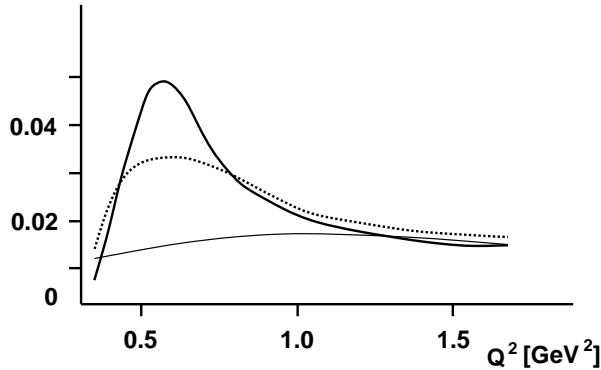}}
\caption{Formation of a ``bump'' on the phenomenological side of sum rule
for $q=1,\,2,$ and $3$. As $q$ increases, the bump becomes more
and more pronounced.}
\label{bump}
\end{figure}

Another argument to support our claim that modulating factors
can extend the validity of sum rules to the intermediate region
is the appearance of ``bumps'' observed numerically 
on the phenomenological side of sum rules in \cite{FK}
at $Q^2\approx 0.5 GeV^2$ (see Fig. \ref{bump}).
%
In the absence of modulating factors,   
it was impossible to match the fits of both sides of sum rule (\ref{fksum})
in the intermediate energy region  and a conclusion 
was drawn that the Borel summability is optimal and that by
any departure from the Borel summability sum rules get worse \cite{FK}.
However, if modulating factors are included, then, since
$G(1/Q^2)\rightarrow 0$ in the limit $Q^2\rightarrow 0$, 
$G(1/Q^2)$ will develop a maximum  at the 
point where $\chi^{(n-1)}(\varepsilon_n/Q^2)$ overcomes polynomial 
increase of $1/Q^{2n}$ (in the limit $Q^2\rightarrow 0$). This 
turning point then provides a way of determining modulating constant
$\varepsilon_n$ phenomenologically.

{\bf 7.} {\em Discussion and conclusions}.-
We have provided a general derivation of sum rules
within the framework of summability methods, without any recourse
to a formal differential operator (\ref{lm-rel}) (cf. \cite{SVZ,FK}).

We have shown that unless symmetries of a 
physical model cause  some particular orders
of the asymptotic expansion (\ref{rex}) of the polarization
operator $\Pi$ to vanish, the only reasonable choices of weight
functions, from the set (\ref{chiab}) and (\ref{chi0})
we have considered, are $e^{-t}$ and $e^{-e^t}$.
The use of  $e^{-e^t}$ instead of $e^{-t}$ in sum rules means
also a significant {\em qualitative change}:
due to very special properties of the summability
method based on the weight function $e^{-e^t}$ \cite{AM},
one is not bound to consider dispersion relation (\ref{disr1})
for only purely imaginary frequencies 
as in the case of $e^{-t}$ but, instead,
one can consider dispersion relation and derive sum rules for any
frequency in the upper-half complex plane. This can have
useful applications in other physical models.
Moreover, a sum rule with $e^{-e^t}$ 
has two advantages over sum rules with $e^{-t}$:
\begin{itemize}
\item  the high-energy part of the integral is cut-off more
effectively;
\item the onset of the cut-off in the integral (\ref{gsr}) starts
sooner, already at $s\sim Q^2 \ln 2$ and not at $s \sim Q^2$
as in the latter case.
\end{itemize}
All the above advantages may compensate for a slower,
$(\ln n)^n/n!$ damping 
[cf. Eq. \ref{chi0d})] of higher orders of expansion (\ref{rex}) 
in comparison to $1/n!$ damping in the case of $e^{-t}$.

Further, we have proved that if a general weight function
is used, the sum rule must contain modulating factors.
The latter provide the low frequency regularization of sum rules
and could extend their validity up to the intermediate region of 
energies. The very existence of the modulating factors in a 
sum rule with weight function different from $e^{-t}$ results from 
the universal $1/t$ decay [see Eq. (\ref{unib})]  of the moment 
function $F(-t)$. It is amazing to note that a modulating factor 
can be also present in the case of the Borel summability method,
and may lead to essential improvement of the fit. One cannot
rule out its presence unless one has absolute control over
numerical values of expansion constants in the asymptotic
behaviour (\ref{rex}) of $\Pi$. If not, one can always
find a positive constant $\varepsilon$ such that replacement
of  $1/Q^2$ by $1/(\varepsilon +Q^2)$ makes a change
in the expansion (\ref{rex}) which is within the error
due to our incomplete knowledge of the expansion constants $a_n$.
Then, the use of (\ref{qn1}) inevitably leads to
a modulating factor. The only exception of the Borel summability method
is that, in principle,
the modulating constants $\varepsilon_n$  can be sent to zero 
and the limit exists
[giving the SVZ sum rule (\ref{srf})],
while for all  other summability methods, the modulating constants 
{\em must} be nonzero.
The modulating factors were not seen using 
the formal infinite-order differential operator $\hat{L}_M$  
involving two limiting procedures [cf. Eq. \ref{lm-rel})], because
the latter set  the argument of the modulating factors 
to the origin rendering them $Q^2$ independent.

One of us (A.M.) would like to thank J. Stern
for raising his interest in sum rules.
J.F. gratefully acknowledges the hospitality of the CERN TH 
Division.
This work was supported by EPSRC grant number GR/J35214.
Partial support by the Grant Agency of the Czech Republic under
Project Nos. 202/93/0689 (A.M.) and
110/93/136 (J.F.) is also gratefully acknowledged.



\begin{thebibliography}{99}
\bibitem{Fc}Functions $\chi(t)$ and $F(t)$ play important roles
in the theory of summability methods. If $f(z)$ has an
(either asymptotic or the Taylor) expansion $f(z)=\sum_n a_nz^n$,
then $f(z)=\int_0^\infty \chi(zt) T(f)(t)\,dt$,
where locally, within its radius convergence,
$T(f)(t)=\sum_n a_n t^n/\mu(n)$. Outside its radius convergence,
analytic continuation $T(f)(t)$ is given by
$T(f)(t)=\oint_{C} F(t/z) f(z) \,dz$, where $C$ is contour specified
by the condition that $\mbox{Re}\ln \,F(t/z)=$ const. See
\cite{Mor} for details.
\bibitem{Abs}We remind the reader that Im $\Pi$ has a definite sign,
Im $\Pi\leq 0$, and the convergence of the dispersion integral
in (\ref{disr}) is equivalent to its absolute convergence.
\bibitem{Uni}Except for cases where $\beta= \alpha$, in which case, provided that
$\alpha$ is not an integer, the leading term in the
asymptotic exansion of $F(-t)$ is of order $t^{-2}$.
If $\alpha$ is an integer and $\alpha\geq 2$, weight function $\chi(t)$
decays as $e^{-\sqrt{t}}$ or slower which is not interesting
for our purposes.
\end{thebibliography}
\end{document}